\documentclass[iop]{emulateapj}

\usepackage{epsfig}
\usepackage{natbib}
\usepackage{graphicx}
\def\simgt{\lower.5ex\hbox{$\; \buildrel > \over \sim \;$}}
\def\simlt{\lower.5ex\hbox{$\; \buildrel < \over \sim \;$}}

\def\amin{\ifmmode^{\prime}\else$^{\prime}$\fi}
\def\asec{\ifmmode^{\prime\prime}\else$^{\prime\prime}$\fi}

\def\simgt{\lower.5ex\hbox{$\; \buildrel > \over \sim \;$}}
\def\simlt{\lower.5ex\hbox{$\; \buildrel < \over \sim \;$}}

\newcommand\chandra{{\it Chandra}}
\newcommand\Chandra{{\it Chandra}}
\newcommand\xmm{{\it XMM-Newton}}

\newcommand\nustar{{\it NuSTAR\/}}

\newcommand\suzaku{{\it Suzaku\/}}

\def\sga{Sgr A*}

\def\sgrAE{G359.89$-$0.08}
\def\snr{G359.92$-$0.09}
\def\mc{M$-$0.13$-$0.08}

\slugcomment{Accepted by The Astrophysical Journal}

\shorttitle{NuSTAR Observation of Sgr~A$-$E}
\shortauthors{S. Zhang et al.}

\begin{document}

\title{High Energy X-ray Detection of \sgrAE~(Sgr~A$-$E):\\ Magnetic Flux Tube Emission Powered by Cosmic Rays?}

\author{Shuo Zhang\altaffilmark{1}, Charles J. Hailey\altaffilmark{1}, Frederick K. Baganoff\altaffilmark{2}, Franz E. Bauer\altaffilmark{3,4}, 
Steven E. Boggs\altaffilmark{5},  William W. Craig\altaffilmark{5,6}, Finn E. Christensen\altaffilmark{7}, Eric V. Gotthelf\altaffilmark{1}, Fiona A. Harrison\altaffilmark{8}, Kaya Mori\altaffilmark{1},  Melania Nynka\altaffilmark{1}, Daniel Stern\altaffilmark{9}, John A. Tomsick\altaffilmark{5} and William W. Zhang\altaffilmark{10}}

\altaffiltext{1}{Columbia Astrophysics Laboratory, Columbia University, New York, NY 10027, USA; shuo@astro.columbia.edu}
\altaffiltext{2}{Kavli Institute for Astrophysics and Space Research, Massachusetts Institute of Technology, Cambridge, MA 02139, USA}
\altaffiltext{3}{Instituto de Astrof\'{\i}sica, Facultad de F\'{i}sica, Pontificia Universidad Cat\'{o}lica de Chile, 306, Santiago 22, Chile}
\altaffiltext{4}{Space Science Institute, 4750 Walnut Street, Suite 205, Boulder, CO 80301, USA}
\altaffiltext{5}{Space Sciences Laboratory, University of California, Berkeley, CA 94720, USA}
\altaffiltext{6}{Lawrence Livermore National Laboratory, Livermore, CA 94550, USA}
\altaffiltext{7}{DTU Space - National Space Institute, Technical University of Denmark, Elektrovej 327, 2800 Lyngby, Denmark}
\altaffiltext{8}{Cahill Center for Astronomy and Astrophysics, California Institute of Technology, Pasadena, CA 91125, USA}
\altaffiltext{9}{Jet Propulsion Laboratory, California Institute of Technology, Pasadena, CA 91109, USA}
\altaffiltext{10}{NASA Goddard Space Flight Center, Greenbelt, MD 20771, USA}

\begin{abstract}

We report the first detection of high-energy X-ray ($E>10$~keV) emission from the Galactic Center non-thermal filament \sgrAE\ (Sgr~A$-$E) using data acquired with the Nuclear Spectroscopic Telescope Array (\nustar). 
The bright filament was detected up to $\sim50$~keV during a \nustar\ Galactic Center monitoring campaign.
The featureless power-law spectrum with a photon index  $\Gamma \approx 2.3$ confirms a non-thermal emission mechanism. 
The observed flux in the $3-79$~keV band is $\rm F_{X} = (2.0 \pm 0.1) \times 10^{-12}$~ erg~cm$^{-2}$~s$^{-1}$, corresponding to an unabsorbed X-ray luminosity $\rm L_{X} = (2.6 \pm 0.8) \times 10^{34}$~erg~s$^{-1}$ assuming a distance of 8.0~kpc. 
Based on theoretical predictions and observations, we conclude that Sgr~A$-$E is unlikely to be a pulsar wind nebula (PWN) or supernova remnant$-$molecular cloud (SNR$-$MC) interaction, as previously hypothesized.
Instead, the emission could be due to a magnetic flux tube which traps TeV electrons.
We propose two possible TeV electron sources: old PWNe (up to $\sim100$~kyr) with low surface brightness and radii up to $\sim 30$~pc or molecular clouds (MCs) illuminated by cosmic rays (CRs) from CR accelerators such as SNRs or \sga.

\end{abstract}

%%%%%%%%%%%%%%%%%%%%%%%%%%%%%%%%%%%%%%%%%%%%%%%%%%%%%%%%%%%%%%%%%%%%
\section{Introduction}

The Galactic Center (GC) hosts not only the supermassive black hole Sagittarius~A* (Sgr~A*), supernova remnants (SNRs), pulsar wind nebulae (PWNe), dense molecular clouds and star clusters, but also many mysterious non-thermal filamentary structures. 
Originally detected at radio wavelengths (e.g. \citealp{Yusef1984}), many non-thermal filaments were later revealed to be strong X-ray emitters (e.g. \citealp{Lu2008, Johnson2009}). 
Within $(l,b) = 1^{\circ} \times 0.\! ^{\circ}5$ of \sga, numerous ($\approx 17$) X-ray filaments are now well-resolved on arcsecond scales in
\chandra\ observations \citep{Lu2008, Muno2008, Johnson2009}.
But their emission mechanism and nature have been under debate since their discovery. 

Among the GC non-thermal filaments, \sgrAE\ (XMM~J17450$-$2904)
\citep{Sakano2003, Lu2003}, the X-ray counterpart to Sgr~A$-$E \citep{Ho1985}, is by far the most luminous. 
Discovered in archival \xmm\ and \chandra\ observations of the GC \citep{Sakano2003, Lu2003}, it was noted for its highly absorbed featureless spectrum and wisp-like linear emission extending $\sim 20^{\prime\prime}$ nearly perpendicular to the Galactic plane.
The large X-ray absorption column is consistent with a GC origin \citep{Sakano2003}.  
The X-ray wisp was identified as a plausible counterpart to a radio filament recorded in archival VLA images of the GC at 2-cm, 6-cm, and 20-cm wavelengths \citep{Ho1985, Yusef1987, Lang1999}.  
The radio spectral index ($\alpha$, $S_\nu \sim \nu^\alpha$) was measured to be -0.4 by \citet{Ho1985} using 2-cm and 6-cm continuum data, and more recently derived as -0.17 by \citet{Yusef2005} using high-resolution continuum data at a number of wavelengths between 2 and 20 cm.
The negative spectral index suggested a non-thermal nature, which was confirmed by detection of radio polarization.
Although their morphologies are similar, the X-ray feature is significantly offset from the radio wisp ($\sim 10^{\prime\prime}$).
The more compact X-ray emission region suggests that the difference between radio and X-ray morphologies could be due to synchrotron cooling losses in an advective flow \citep{Yusef2005, Sakano2003}.

One possibility for the origin of the synchrotron particles is a ram-pressure confined PWN as proposed by \citet{Lu2003}. 
They reported the marginal (2.5~$\sigma$) detection of a point source (CXOU~J174539.6$-$290413), which they speculated to be the pulsar that powers the PWN.
However, the point source was not confirmed by deeper \chandra\ observations \citep{Yusef2005, Johnson2009}. 
The PWN scenario also predicts spectral steepening towards the pulsar, which was investigated by \citet{Johnson2009} using deep \chandra\ observations.
Their detailed spatially resolved spectral analysis showed no appreciable spectral steepening across either the minor or major axis of the filament, to within the 90\% confidence error bar, thus disfavoring a PWN scenario.

Another plausible explanation is a supernova remnant and molecular cloud (SNR$-$MC) interaction, in which the Sgr~A$-$E radio emission is due to the interaction between the shock front of a second SNR south of Sgr~A East and the molecular cloud \mc, also known as the 20~km~${\rm s}^{-1}$ cloud \citep{Ho1985, Coil2000, Yusef2005}. 
The second SNR, \snr, is believed to explain the circular feature south of Sgr A East in the 20-cm continuum emission map, which is shown in \citet{Pedlar1989}. 
Because of the observed redshifted gas at the position of Sgr~A$-$E, \citet{Coil2000} suggested that Sgr~A$-$E is the result of a SNR shock wave expanding into the 20~km~${\rm s}^{-1}$ cloud behind it along the line-of-sight. 
This is contradicted by the very high absorption derived in X-ray observations, which suggests Sgr~A$-$E is embedded or behind the 20~km~${\rm s}^{-1}$ cloud \citep{Johnson2009}. 
Because of these controversies, no compelling conclusions have been drawn to date about the nature of Sgr~A$-$E.

In this paper we report the first detections of hard X-ray emission from Sgr~A$-$E up to $\sim50$ keV. 
In \S 2 we present the observations and data reduction, while in \S 3 we discuss the morphology, and in \S 4 we discuss the spectroscopy. 
\S 5 reports on a pulsation search. Three possible scenarios explaining Sgr~A$-$E emission are discussed in \S 6. Finally we present our conclusions in \S 7.

%%%%%%%%%%%%%%%%%%%%%%%%%%%%%%%%%%%%%%%%%%%%%%%%%%%%%%%%%%%%%%%%%%%%
\section{NuSTAR observations}

\nustar\ is the first in-orbit focusing telescope operating in the broad X-ray energy band from 3 to 79 keV \citep{Harrison2013}. 
Sgr~A$-$E is in the GC field, which has been monitored by \nustar\ since July 2012.
In all the observations, the GC region was imaged with the two co-aligned X-ray optics/detector pairs, focal plane modules FPMA and FPMB, providing an angular resolution of $58\asec$ Half Power Diameter (HPD) and $18\asec$ Full Width Half Max (FWHM) over the $3-79$~keV X-ray band, with a characteristic spectral resolution of 400~eV (FWHM) at 10~keV. 
The nominal reconstructed \nustar\ astrometry is accurate to $8\asec$ (90\% confidence level) \citep{Harrison2013}, but improves significantly after image registration ($\sim 2 \asec$).

During the \nustar\ GC monitoring campaign, three observations were centered on Sgr~A*, and six observations were conducted in 2012 as part of a broader GC survey. 
In addition, \nustar\ triggered ToO observations of the newly discovered magnetar SGR~J1745$-$29 near Sgr~A* in 2013 (\citealp{Mori2013}, Kaspi et al. submitted). 
Sgr~A$-$E was fully captured in six observations, listed in Table \ref{tab:obs}, resulting in a total exposure time of 338.5~ks. 
We analyzed all the data sets for imaging, spectral and timing information. 
The data were reduced and analyzed using the \nustar\ {\it Data Analysis Software NuSTARDAS} v.1.1.1. and HEASOFT v. 6.13, and filtered for periods of high instrumental background due to South Atlantic Anomaly (SAA) passages and known bad detector pixels. 
Photon arrival times were corrected for on-board clock drift and precessed to the Solar System barycenter using the JPL-DE200 ephemeris and the coordinates of the \chandra\ peak emission of the \sgrAE\ at RA=$17^{h}45^{m}40^{s}.4$, Dec=$-29^{\circ}04^{\prime}29\farcs0$ (J2000.0) \citep{Lu2003}.

\begin{deluxetable}{lccccc}                                                                                                                
\tablecaption{NuSTAR observations of Sgr~A-E.}
\tablewidth{0pt}
\tablecolumns{4}                                                                                                                    
\tablehead{ \colhead{Observation}   &   \colhead{Start Date}   &   \colhead{Exposure}     &     \colhead{Target} \\
\colhead{ID}  & \colhead{(UTC)}  &  \colhead{(ks)}   & \colhead{ } }  
\startdata
30001002001  & 2012 07 20  & 154.22  & Sgr~A*\\
30001002004  & 2012 10 16  & 49.56  & Sgr~A*\\
40010001002  & 2012 10 13  & 23.91  & GC Survey\\
40010002001  & 2012 10 13  & 24.22  & GC Survey\\
30001002006  & 2013 04 27  & 36.99  & Magnetar ToO \\
80002013002  & 2013 04 27  & 49.60  & ~Magnetar ToO
\enddata
\label{tab:obs}
\end{deluxetable}

%%%%%%%%%%%%%%%%%%%%%%%%%%%%%%%%%%%%%%%%%%%%%%%%%%%%%%%%%%%%%%%%%%%%%%%
\section{Morphology}

\begin{figure*} [t] 
\centerline{ \hfill
    \psfig{figure=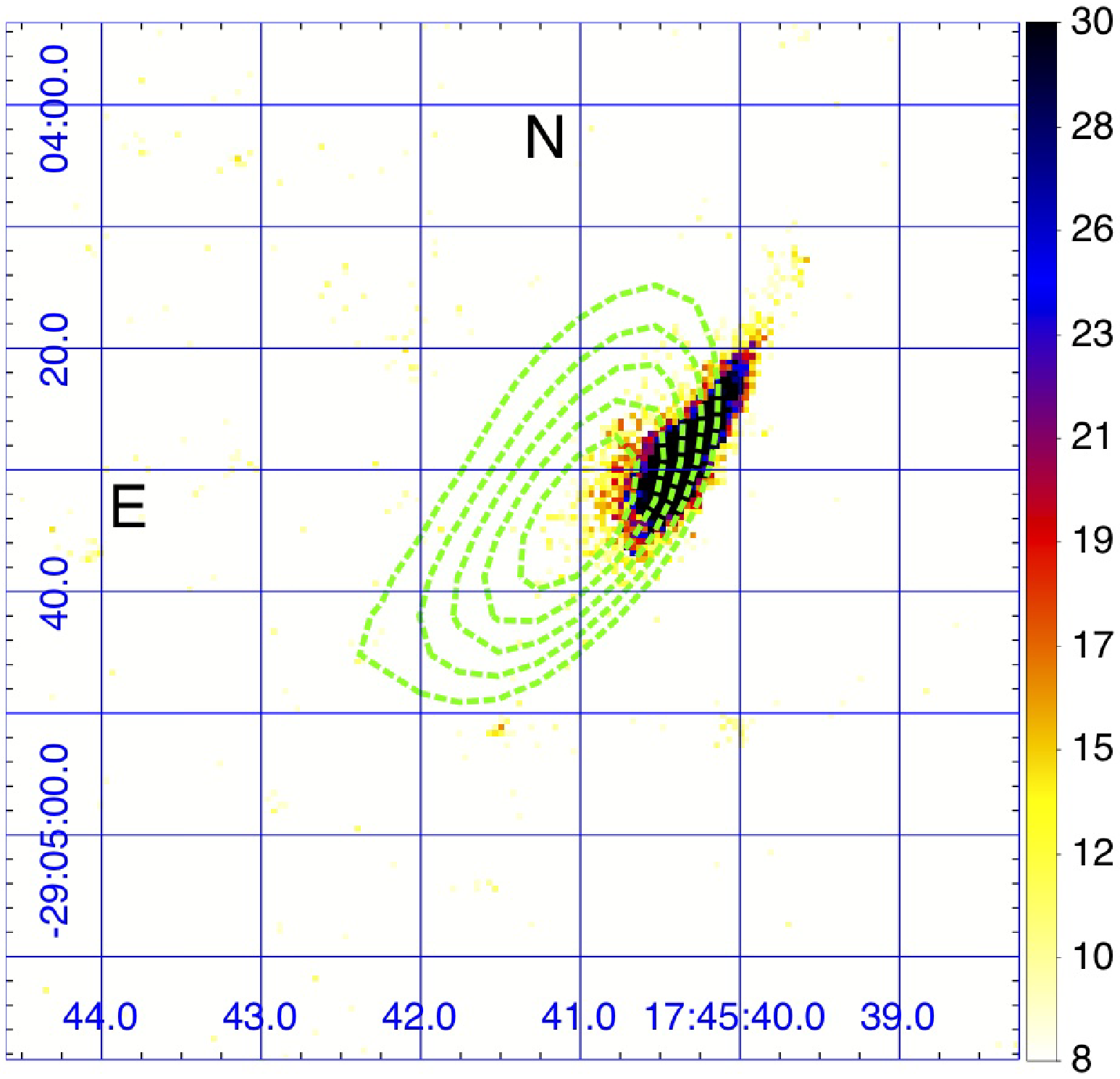,width=0.5\linewidth, height=0.4\linewidth} \hfill
    \psfig{figure=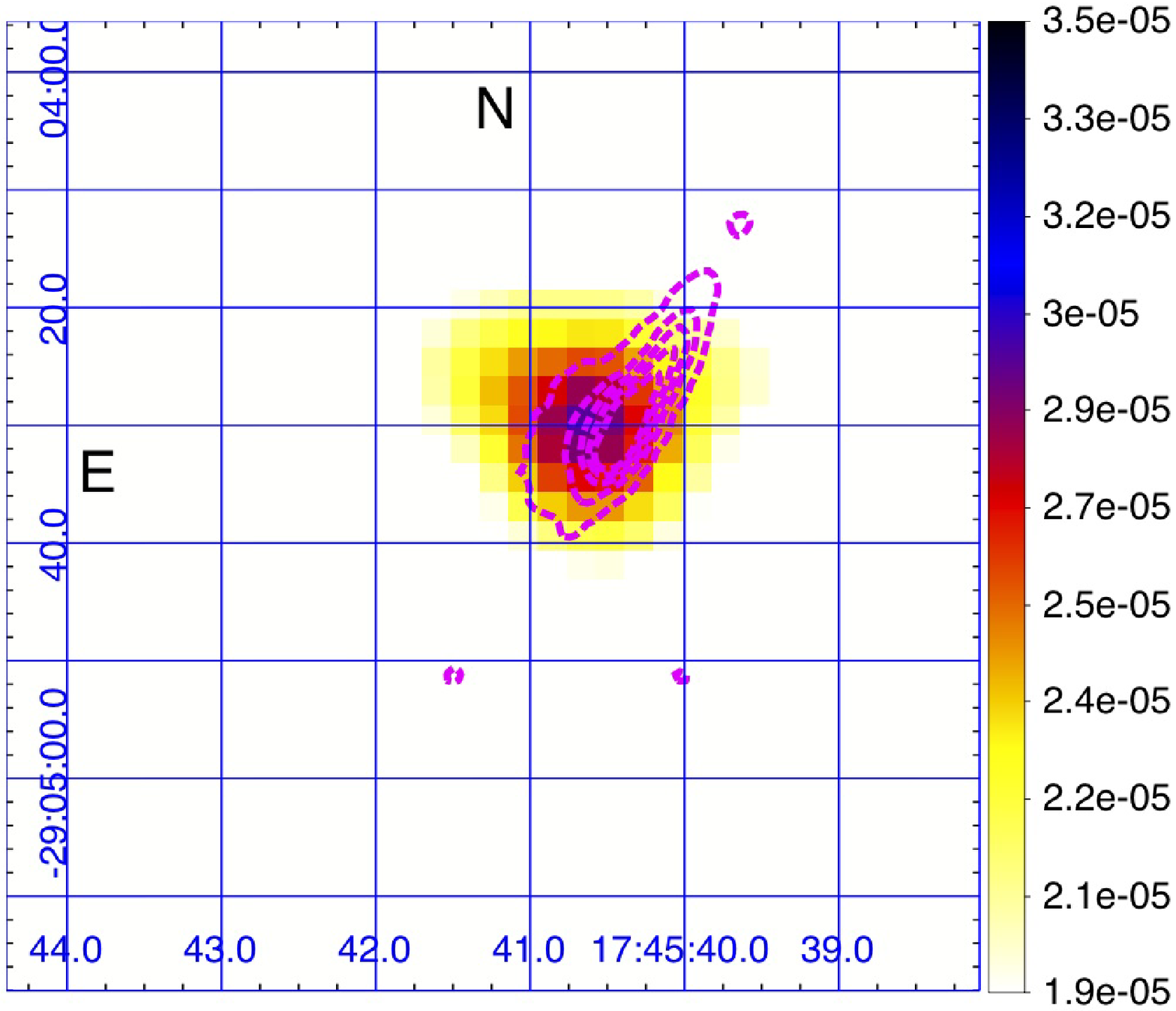,width=0.5\linewidth, height=0.4\linewidth} \hfill  }
\caption{
Left panel: \chandra\ 2$-$10~keV image overlaid with VLA 20-cm continuum contours (green dashed) of Sgr~A$-$E. The X-ray feature is $\sim 10 \asec$ offset from the radio wisp. Right panel: \nustar\ 10$-$50~keV mosaic image overlaid with \chandra\ 2$-$10~keV contours (magenta dashed) of Sgr~A$-$E. The image is shown in a linear color scale and the scale range chosen to highlight the high energy centroid. The 10$-$50~keV emission is consistent with a point source, and its centroid is consistent with the 2$-$10~keV emission centroid, $\sim 20 \asec$ southeast of the putative pulsar.
}
\label{fig:img}
\end{figure*}

We made \nustar\ mosaiced images to illustrate the morphology of the Sgr A$-$E region based on the following steps.
We first registered images using bright sources available in individual observations \citep{Nynka2013}.
The resulting offsets were used to correct narrow band images which were exposure-corrected and combined.
Because the $\sim20\asec$ elongation perpendicular to the Galactic plane is not resolved by \nustar\ below 10~keV, we also made \chandra\ mosaiced images to illustrate the wisp-like shape of Sgr~A$-$E.
The \chandra\ image was made from all the archived \chandra\ data between 1999-09-21 and 2012-10-31 available for Sgr~A$-$E.
Individual observations were registered to a common astrometric frame and merged.
In total, the resulting \chandra\ image includes $\approx$1.8\,Ms and $\approx$3.4\,Ms of ACIS-I and HETG 0th order data, respectively.
To compare with the radio morphology, we made 20-cm continuum contours of Sgr~A$-$E out of the VLA 20-cm continuum map from \citet{Yusef1984}.
Figure \ref{fig:img} (right panel) shows the $10-50$~keV \nustar\ mosaic overlaid with the \chandra\ $2-10$~keV contours. 
Detection of Sgr~A$-$E in the 10$-$50~keV energy band is consistent with a point source.
The high energy  ($>10$~keV) centroid lies closer to the south-east end of the filament, consistent with the position of the low energy ($<10$~keV) centroid.
in Figure \ref{fig:img} (left panel) we show the \chandra\ 2$-$10~keV image overlaid with the VLA 20-cm contours to illustrate the filament shape and the $\sim10\asec$ offset between the radio and X-ray emission.

%%%%%%%%%%%%%%%%%%%%%%%%%%%%%%%%%%%%%%%%%%%%%%%%%%%%%%%%%%%%%%%%%%%%%%%%
\section{Spectroscopy}

We analyzed the full spectral data from the six observations
using an extraction region of 60\asec\ in radius centered on Sgr~A$-$E. 
Local background was extracted from individual observations.
We joint-fitted the data with \xmm\ observations to better constrain the column density. 
Two \xmm\ (PN, MOS1 and MOS2) observations (obsID 0658600101 and 0658600201) were used, yielding a 102.5 ks exposure time in total. 
The data were proceeded with \xmm\ Scientific Analysis System {\tt SAS} version 13.0.0.
A 40\asec\ radius aperture was used to extract source photons, and the background spectra were extracted from local surrounding regions.
Joint spectral analysis was done in the $1-12$~keV band for \xmm\ and $5-79$~keV band for \nustar\ data using {\tt XSPEC} version 12.8.0 \citep{Arnaud1996}.

\begin{figure}

\psfig{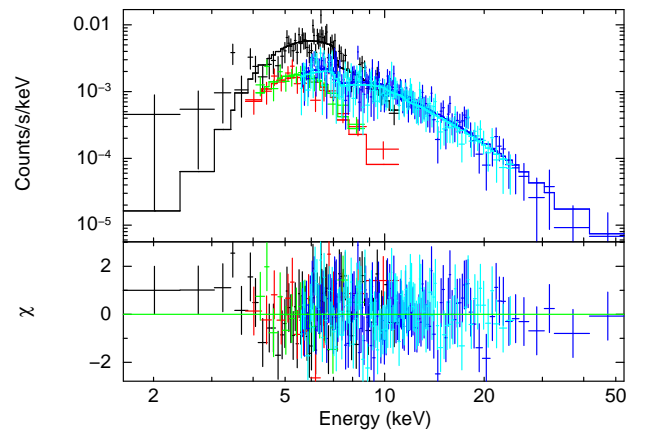}

\caption{\nustar\ (blue and cyan for FPMA and FPMB, respectively), and \xmm\ (black, red and green for PN, MOS1, MOS2, respectively) spectra jointly fitted to an absorbed power-law model. The crosses show the data points with 1-$\sigma$ error bars, and the solid lines show the best fit model. The lower \
panel shows the deviation from the model in units of standard deviation.}
\label{fig:spec}
\end{figure}

The spectra up to $\sim50$~keV are well-fit ($\chi^2_{\nu} =0.91$ for 298 DoF) by a simple absorbed power-law model with  photon index
$\Gamma=2.28^{+0.17}_{-0.18}$ and $N_{\rm H} = (7.2 \pm 1.0) \times10^{23}~{\rm cm}^{-2}$, using the {\tt Tbabs} absorption model with \citet{Verner96} atomic cross sections and \citet{Wilms2000} abundances (see Table \ref{tab:specfit}).
The updated abundances increase the derived absorption column density by a factor of two compared to previous measurements \citep{Sakano2003, Lu2003, Yusef2005, Johnson2009}. 
The high column density supports Sakano's argument that the source is embedded or behind the  20 km~${\rm s}^{-1}$ cloud \mc. 
The $3-79$~keV  flux is $\rm F_{X} = (2.0 \pm 0.1) \times 10^{-12}$~ erg~cm$^{-2}$~s$^{-1}$, corresponding to a luminosity $\rm L_{X} = (2.6 \pm 0.8) \times 10^{34}$~erg~s$^{-1}$ at 8.0~kpc.
The fitting result is consistent with previous measurements \citep{Sakano2003, Lu2003, Yusef2005}, while the photon index is much better constrained.
The featureless spectra (Figure \ref{fig:spec}) also demonstrate a proper background subtraction, lacking the 6.7~keV line from the GC diffuse emission.
This agrees with \chandra\ and \xmm\ measurements, where likewise no line features were detected.

\begin{deluxetable}{lcccc}                                                                                                               
\tablecaption{Power-law model of the {\it XMM} and {\it NuSTAR} data.}
\tablecolumns{2}                                                                                                                    
\tablehead{ \colhead{Parameter}   &  \colhead{Value}  }
\startdata
$N_{\rm H}$ (10$^{23}$ cm$^{-2}$)   & $7.2\pm1.0$ \\
$\Gamma$                                          &  $2.28^{+0.17}_{-0.18}$ \\
flux (erg/cm$^2$/s)                            & $(2.0 \pm 0.1) \times 10^{-12}$ \\
$\chi^2_{\rm \nu}$ (DoF)                         & 0.91 (298) \\
\enddata
\tablecomments{$N_{\rm H}$ is the column density,  $\Gamma$ is the photon index of the power-law. 
The $3-79$ keV flux is given. The goodness of fit is evaluated by the reduced $\chi^2$ and the degrees of freedom is given in parentheses. 
The errors are 90\% confidence.}
\label{tab:specfit}
\end{deluxetable}

%%%%%%%%%%%%%%%%%%%%%%%%%%%%%%%%%%%%%%%%%%%%%%%%%%%%%%%%%%%%%%%%%%%
\section{Pulsation Search}

Although the lack of evidence for a point source in the \chandra\
energy band argues against a pulsar powering \sgrAE, we nevertheless
searched the unexplored \nustar\ data above 10~keV for a coherent
signal.  The high time-resolution of the \nustar\ data allows a search
for pulsations with $P \geq 4$~ms, covering the expected range for an
isolated rotation-powered pulsar. For each observation listed in
Table~\ref{tab:obs} we generated light curves by extracting photons in
the $10-30$~keV range from an $18''$ radius aperture centered on the
source, to optimize the signal-to-noise ratio.  We searched each
light-curve for significant power from a coherent signal using a fast Fourier transform (FFT)
sampled at the Nyquist frequency. To account for possibly significant
spin-down of a highly energetic pulsar during the observation span of
the longest observations (ObsID 30001002001), we performed an
``accelerated'' FFT search. This required four frequency derivative steps
to be sensitive to $\dot E_{\rm max} = 10^{38}$~erg~s$^{-1}$.

From a search of all the observations, the most significant signal has
a power of $38.12$ for ObsID 30001002001, corresponding to a
probability of false detection of $\wp > 1$ for $4 \times 2^{28}$ search
trials. The resulting period is not reproduced in the other
observations. We conclude that no pulsed X-ray signal is detected in
the $>10$~keV band from \sgrAE. After taking into account the local
background, estimated from a annulus region around the source, we
place an upper limit on the pulse fraction at the 99.73\% confidence
level ($3\sigma$) of $f_p < 66\%$ for a blind search for a sinusoidal
signal $P>4$~ms.

%%%%%%%%%%%%%%%%%%%%%%%%%%%%%%%%%%%%%%%%%%%%%%%%%%%%%%%%%%%%%%%%%%%
\section{Discussion}

\subsection{PWN Scenario}

The featureless power-law spectrum extending up to $\sim50$~keV is consistent with synchrotron emission.  
Using the most recent measurement of Sgr A$-$E radio spectral index (-0.17, \citealp{Yusef2005}), the steepening in the spectral index is $\sim1$, consistent with the synchrotron picture suggested by \citet{Sakano2003} and \citet{Yusef2005}.
Assuming a magnetic field of 100~$\mu$G~$-$~300~$\mu$G as estimated by \citet{Yusef2005} and \citet{Ho1985} respectively, $\lesssim100-200$~TeV electrons are required to generate up to 50 keV synchrotron emission.
The flux of Sgr~A$-$E has maintained the same level from 2003 to 2013; thus, there must be continuous injection of relativistic electrons considering the $\sim 2-6$~yr cooling lifetime of $\sim 100-200$~TeV electrons emitting hard X-rays through synchrotron radiation.
One explanation for the origin of the required high-energy electrons is the PWN picture proposed by \citet{Lu2003}.
In this scenario, the putative pulsar detected with a signal-to-noise ratio of $\sim2.5$ is moving north-west supersonically, generating the X-ray tail behind it to south-east.
The authors also pointed out that the $\sim10\asec$ offset between the radio and X-ray emission (Figure \ref{fig:img}) can be explained by a ram-pressured confined PWN, because the radio emission comes from accumulated radio synchrotron particles (with longer lifetimes than X-ray synchrotron particles) over a longer history of the PWN, in a region close to where the pulsar was born and offset from the X-ray feature.  
However, two X-ray observation results conflict with this scenario. 
First, the point source interpreted as the pulsar was not detected in deeper \chandra\ observations \citep{Yusef2005, Johnson2009}. 
Second, the centroid of the higher energy emission ($>$10 keV) sits close to the southeast end of the filament, $\sim 20 \asec$($\sim 0.8~\rm pc$ at 8.0~kpc) from the putative point source.
If the point source is indeed a pulsar powering the PWN, the $\sim100$~TeV electrons in the post-shock outflow can only travel up to $\sim0.05~\rm pc$ given the post-shock speed of $\sim 0.1 \rm c$ \citep{Kennel1984} before losing most of their energy through synchrotron emission. 
Thus, the hard X-ray emission should be produced in the vicinity of the termination shock around the pulsar, which is not consistent with observations.
Thus, both the \nustar\ hard X-ray observations and the deep \chandra\ observations argue against the PWN picture with a pulsar moving northwest.

But can Sgr~A$-$E be a PWN moving southeast?
Although it is consistent with the fact that the hard X-ray centroid sits close to the southeast end of the PWN, there is no PWN with radio emission leading the X-ray head based on investigations of PWNe in the catalogue of \citet{KP2008}.
Non-detection of a pulsar from both the image and the timing analysis also does not support the PWN scenario.

Another powerful argument against the PWN picture comes from the radio morphology of Sgr~A$-$E.
The 20-cm continuum map (Fig. 21, Fig. 22 in \citealt{Yusef2004}) shows two long and highly curved filamentary structures, Sgr~A$-$E and Sgr~A$-$F.  
Sgr~A$-$F is not detected by \nustar\ since its X-ray flux is about two orders of magnitude lower than that of Sgr~A$-$E \citep{Yusef2005}, below the \nustar\ detection threshold.
A new 6-cm continuum radio map made with JVLA (B and C arrays) shows sub-arcsecond structures in the radio filaments in unprecedented detail and reveals that both Sgr~A$-$E and Sgr~A$-$F consist of a bundle of bright radio filaments that are part of a large-scale filamentary structure extending north to Sgr~A~East (M. Morris, private communication).
Based on all these results, we suggest that Sgr~A$-$E is unlikely to be a PWN.

\subsection{SNR$-$MC Interaction}        

Another possible explanation of the Sgr~A$-$E emission mentioned by several authors is the shock wave front of an SNR driving through the 20~km~${\rm s}^{-1}$ cloud. 

Having derived a  power-law spectrum with photon index of $2 \pm 0.5$ from \Chandra\ data, \citet{Yusef2005} suggested that particles accelerated to relativistic energies emitting X-ray synchrotron emission can be explained by \citet{Bykov2000}, in which a SNR forward shock wave propagates in a molecular cloud, producing non-thermal electrons.
In this model, non-thermal emission in the $1-100$~keV energy band comes from the low-energy tail of the non-thermal Bremsstrahlung and inverse Compton scattering (IC scattering) peaking at $\sim \rm GeV$ energies, thus producing a sharply rising $\nu \rm F_{\nu}$ spectrum (photon index $\Gamma \le 1.5$) in the X-ray band. 
However, the broadband \nustar\ and \xmm\ spectra constrain the photon index to $2.28^{+0.17}_{-0.18}$, which cannot be explained by the Bykov model.

In a more recently developed SNR$-$MC interaction model by \citet{Tang2011}, the X-ray emission comes from both primary particles and secondary electron-positron pairs produced via $p-p$ interactions in the shell evolving in the interstellar medium (ISM) and in the shell interacting with the molecular cloud. 
If the shell evolves in the molecular cloud, Bremsstrahlung and IC scattering contribute to the X-ray emission, also predicting very hard spectra similar to the Bykov model.
Moreover, according to their spectral energy distribution (SED) calculation, the emission from the SNR shell evolving into the ISM is more luminous in X-rays than the shell evolving in the molecular cloud, which is not consistent with the assumption that Sgr~A$-$E is due to the SNR shell driving through the 20~km~${\rm s}^{-1}$ cloud.
Thus this model cannot explain the Sgr~A$-$E spectra or morphology.

Current SNR$-$MC interaction theories cannot explain the X-ray morphology or the spectra with $\Gamma > 2$.  
There is no observational evidence of shock excitation such as OH 1720 MHz masers. 
Further, \snr\ is not even a confirmed SNR, but only speculated to be a SNR based on a circular feature south of Sgr~A East in the 20-cm continuum emission \citep{Ho1985}.
Since there is little supporting evidence for this scenario, we conclude that Sgr~A$-$E is unlikely to be due to an SNR$-$MC interaction.

\subsection{Magnetic Flux Tube}

Based on their radio morphologies, it has been pointed out that non-thermal filaments might trace the GC magnetic field lines (e.g.~\citealp{Yusef1984, Tsuboi1986}).
Their filamentary structures might be magnetic flux tubes, where relativistic electrons get trapped in locally enhanced magnetic fields \citep{Boldyrev2006} and generate synchrotron emission. 
Particularly for Sgr~A$-$E, the radio polarization detection suggests that the local magnetic field lines are parallel to the filament \citep{Ho1985}, which is consistent with this picture.
The Sgr~A$-$E radio structure, a bundle of filaments revealed by the new 6-cm continuum map, also supports the magnetic flux tube interpretation.
The more compact X-ray region compared to the radio wisp region, and the point-like X-ray emission above 10~keV compared to the elongated feature below 10~keV can be explained by synchrotron cooling losses. 
The offset between radio and X-ray emission could be due to differing spatial distributions of GeV and TeV electrons.

A persistent problem with the magnetic flux tube hypothesis has been the origin of the high energy electrons.
Magnetic reconnection zones formed in collisions between magnetic flux tubes and molecular clouds have been proposed as a mechanism for accelerating electrons to high energies (e.g.~\citealp{Lieb2004}). 
\citet{Linden2011} summarized problems with this theory, one of which is that collisional reconnection results in a maximum electron energy of less than 10~MeV, insufficient to produce the observed X-rays by synchrotron radiation. 

We propose two possible high energy electron sources.
\citet{Bamba2010} reported \suzaku\ observations of old PWNe with ages up to $\sim100$~kyr and radii up to $\sim 20-30$~pc.
They showed that the X-ray sizes of the PWNe increase with the characteristic age of the host pulsar. 
In order to explain the observed correlation between the extended X-ray emission and pulsar age, they noted that the magnetic field must decrease with time.
When the PWN magnetic field strength decays to a few $\mu$G (comparable to the GC magnetic field strength), TeV electrons can diffuse up to a few tens of pc with enough energy to emit synchrotron X-rays.
No such old extended PWNe have been observed near the GC, which could be due to the strong GC diffuse emission.
Taking the PWN associated with the 51 kyr old pulsar PSR~J1809$-$1917 for example, the observed PWN size is $21\pm8$~pc \citep{Bamba2010}.
The surface brightness of its large-scale extended emission is $\sim 4 \times 10^{-14}~\rm ergs~cm^{-2}~s^{-1} arcmin^{-2}$ in 0.8$-$7~keV \citep{Kargaltsev2007}, an order of magnitude lower than the GC diffuse emission surface brightness of $\sim (1-4) \times 10^{-13} ~\rm ergs~cm^{-2}~s^{-1} arcmin^{-2}$ in the same energy band \citep{Muno2004}.
If such low surface brightness PWNe exist in the GC, they are very hard to resolve from the GC diffuse emission.
However, if the relativistic electrons were to get trapped in locally enhanced magnetic fields, the synchrotron emission would be enhanced.
With an electron spectral index $\rm p = 2\Gamma-1 \sim 3$, synchrotron emissivity is proportional to $\rm B^{(p+1)/2} \sim B^{2}$.
Thus, when the magnetic field strength increases from the large-scale GC magnetic field of $\sim 10~\mu$G (e.g. \citealp{Tsuboi1985}) to the local Sgr A$-$E magnetic field of $\sim 100-300~\mu$G (\citealp{Yusef2005, Ho1985}), the synchrotron emission should be enhanced by a factor of $\sim 100-900$, i.e. in the case of PWN around PSR~J1809$-$1917, its surface brightness is enhanced to $\sim 0.4-4 \times 10^{-11}~\rm ergs~cm^{-2}~s^{-1} arcmin^{-2}$, significantly higher than the GC diffuse emission.
Thus, old extended PWNe near the GC could serve as TeV electron sources for magnetic flux tubes.
 Besides Sgr~A$-$E, there are several fainter non-thermal filaments detected by \nustar\ above 10~keV, which could potentially be explained by this scenario as well.

Another possible TeV electron source is a molecular cloud illuminated by cosmic rays (CRs) from nearby CR accelerators such as SNRs or \sga\ \citep{Aharonian2005, Aharonian2006}.
CR protons that reach the molecular cloud produce secondary electrons inside the cloud.
The secondary electrons with particle energy between $\sim 100$~MeV and $\sim 100$~TeV can quickly escape the cloud because their diffuse propagation timescale  to escape the cloud is shorter than the energy loss timescale \citep{Gabici2009}.
Current theoretical models do not specifically predict the flux of the electrons escaping from the 20~km~${\rm s}^{-1}$ cloud.
Extending the models to this scenario would be informative.
However, there are phenomenological predictions of such a scenario.
In particular, we expect correlations between hard X-ray emission and magnetic flux tubes associated with MCs.
Indeed there is some evidence of such a correlation from the preliminary analysis of \nustar\ GC survey data (Hailey et al. in prep.).

\section{Summary}
The \nustar~detection of Sgr~A$-$E up to $\sim50$~keV is the first X-ray detection of a non-thermal filament at $>$ 10 keV. 
The featureless power-law model with a photon index of $2.28^{+0.17}_{-0.18}$ confirms a non-thermal emission mechanism.
We present three possible scenarios, a PWN, SNR$-$MC interactions and a magnetic flux tube. 
We conclude that Sgr~A$-$E is unlikely to be a PWN based on its radio and X-ray morphology.
The observations cannot be explained by SNR$-$MC interaction theories.
Thus we propose Sgr~A$-$E could be a magnetic flux tube which traps TeV electrons from old extended PWNe or nearby molecular clouds illuminated by cosmic rays accelerators like SNRs or \sga.
Finally, several fainter non-thermal filaments are also detected above 10~keV by \nustar, showing some evidence of a correlation between hard X-rays and molecular clouds.

\acknowledgements
This work was supported under NASA Contract No. NNG08FD60C, and made use of data from the \nustar\ mission, a project led by the California Institute of Technology, managed by the Jet Propulsion Laboratory, and funded by the National Aeronautics and Space Administration. We thank the \nustar\ Operations, Software and Calibration teams for support with the execution and analysis of these observations. This research has made use of the \nustar\ Data Analysis Software (NuSTARDAS) jointly developed by the ASI Science Data Center (ASDC, Italy) and the California Institute of Technology (USA). SZ is partially supported by NASA Headquarters under the NASA Earth and Space Science Fellowship Program - Grant ``NNX13AM31''.The authors wish to thank Mark Morris for allowing them to view the recently acquired JVLA 6-cm continuum radio map of Sgr~A$-$E.

\end{document}